\documentclass[runningheads]{llncs}
\usepackage{graphicx}
\usepackage{booktabs}
\usepackage[dvipsnames]{xcolor}
\usepackage{arydshln}
\usepackage{graphicx}
\usepackage{comment}
\usepackage{amsmath,amssymb} %
\usepackage{color}
\usepackage{epsfig}
\usepackage{graphicx}
\usepackage[noend]{algpseudocode}
\usepackage{algorithm}
\usepackage{algorithmicx}
\usepackage{multirow}
\usepackage{verbatim}
\usepackage{dblfloatfix}
\usepackage{wrapfig}
\usepackage{enumitem}
\usepackage{adjustbox}

\newcommand{\etal}{\textit{et al}.}
\DeclareMathOperator*{\argmax}{argmax}

\begin{document}
\title{Enhanced MRI Reconstruction Network using Neural Architecture Search}
\titlerunning{EMR-NAS}

\author{Qiaoying Huang\inst{1} \and
Dong Yang\inst{2}\and
Yikun Xian\inst{1}\and
Pengxiang Wu\inst{1}\and
Jingru Yi\inst{1}\and
Hui Qu\inst{1}\and
Dimitris Metaxas\inst{1}}

\institute{
Rutgers University, Department of Computer Science, Piscataway, NJ, USA \and
NVIDIA, Bethesda, MD, USA\\
}

\maketitle              %
\begin{abstract}
The accurate reconstruction of under-sampled magnetic resonance imaging (MRI) data using modern deep learning technology, requires significant effort to design the necessary complex neural network architectures.
The cascaded network architecture for MRI reconstruction has been widely used, while it suffers from the ``vanishing gradient'' problem when the network becomes deep. 
In addition, the homogeneous architecture degrades the representation capacity of the network.
In this work, we present an enhanced MRI reconstruction network using a residual in residual basic block.
For each cell in the basic block, we use the differentiable neural architecture search (NAS) technique to automatically choose the optimal operation among eight variants of the dense block.
This new heterogeneous network is evaluated on two publicly available datasets and outperforms all current state-of-the-art methods, which demonstrates the effectiveness of our proposed method.

\end{abstract}

\section{Introduction}
Magnetic resonance imaging (MRI) is widely used in many clinical applications. 
However, acquiring a fully-sampled MRI scan is time consuming, which is expensive and often uncomfortable to the patient.
In clinical practice, MR data are often undersampled in the Fourier domain to speed up the acquisition process. 
Many researchers have focused on developing new methods to accelerate MRI reconstruction, including a series of compressed sensing methods~\cite{lustig2007sparse,lustig2008compressed,yang2010fast,huang2011efficient,yang2016sparse} and deep learning-based methods~\cite{schlemper2017deep,hyun2018deep,sun2018compressed,zheng2019cascaded,huang2019fr,huang2019mri,pmlr-v102-huang19a,huang2019brain}.  %

\begin{figure}[!t]
    \centering
    \includegraphics[width=\linewidth]{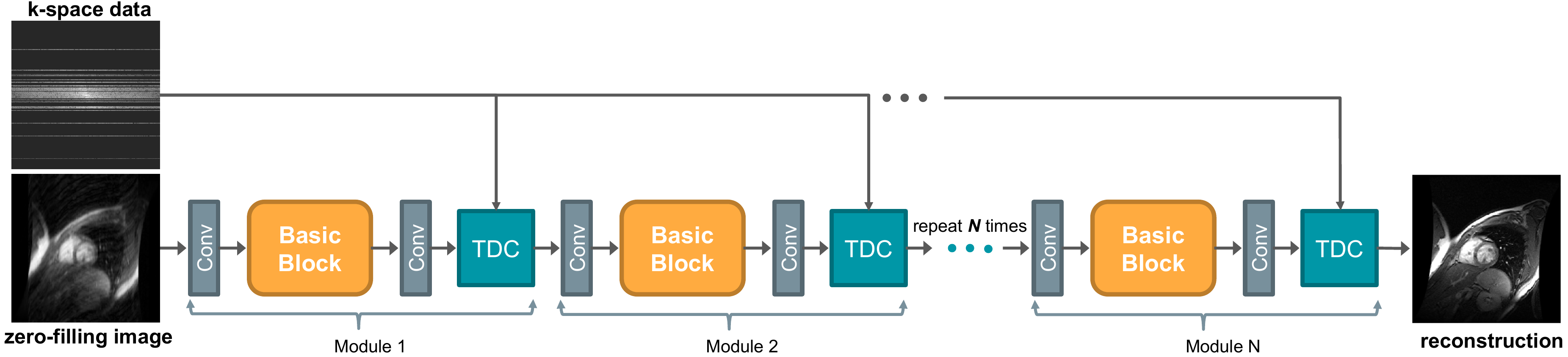}
    \caption{The common network architecture of MRI reconstruction. %
    }
    \label{fig:basic}
\end{figure}

Recently, deep learning-based methods have achieved promising high-quality image reconstruction results.
These methods use a similar framework, as shown in Fig.~\ref{fig:basic}, by stacking the same modules to form a very deep network to directly map the undersampled data to fully-sampled data. 
For example, Schlemper \etal~\cite{schlemper2017deep} propose a deep neural network using cascaded convolutional layers with data consistency (DC) layers to compensate the reconstructed data with the original $k$-space data. 
A UNet combined with DC layers has been shown to achieve good results in MRI reconstruction~\cite{hyun2018deep}.
Sun \etal~\cite{sun2018compressed} propose a recursive dilated network (RDN) and prove that dilated convolution in each recursive block can aggregate multi-scale information within the MRI.
The most recent work~\cite{zheng2019cascaded} uses repeated dilated dense blocks in the framework and improves the DC layer via a two-step compensation in both $k$-space and image domains.
The common feature of these works is to employ a very deep architecture with homogeneous computing blocks.
However, as the depth of the network increases, the model may suffer from the gradient vanishing problem. 
Besides, the homogeneous blocks may limit the feature representation capacity of the network. 

To the end, we propose a deep neural model called \textbf{EMR-NAS} featured by residual-in-residual (RIR) structure and heterogeneous blocks.
The RIR structure \cite{wang2018esrgan} is shown to be effective in alleviating gradient vanishing in tasks such as super resolution.
For heterogeneity, we design various candidate operations inside blocks. To avoid huge manual effort of tuning the best composition of operations, we employ neural architecture search (NAS) technique to automatically decide which one is optimal to improve the ability of feature learning.
NAS achieves promising performance in classification tasks but is seldom explored in the MRI reconstuction domain. 
It is ideal for those tasks that need arduous architecture design, such as in MRI reconstruction.
NAS methods can be separated into two different types: optimizing by reinforcement learning algorithm or by being differentiable with the use of back-propagation.
The differentiable ones are more effective and cost less computational resource.
They alternatively train the shared weights of the network~\cite{pham2018efficient,bender2019understanding} and parameters of the architecture design.
For example, DARTS~\cite{liu2018darts} propose a continuous relaxation of the architecture parameters by a softmax function, allowing an efficient search of the architecture using gradient descent. 
However, it still depends on very large GPU memory and needs a long training time.
Therefore,  they can only search the architecture on a smaller dataset and then transfer it to the a large dataset.
ProxylessNAS~\cite{cai2018proxylessnas} aims to overcome this limitation, by proposing a binarization strategy that activates only specific paths during training to decrease training memory and time dramatically. 
Since this NAS technique can replace human efforts, we apply it in the MRI reconstruction problem to automatically choose the optimal block for boosting the performance. 

Our major contributions are listed below:
(1) To the best of our knowledge, this is the first work to study Neural Architecture Search techniques for MRI reconstruction.
(2) We propose to use a residual in residual (RIR) basic block for the deep reconstruction network. 
(3) We design a new search space with eight novel cell-level operations adapted to MRI reconstruction and they are placed in the RIR basic block to boost the network capacity and performance.
(4) The searched heterogeneous architecture achieves superior performance over current state-of-the-art reconstruction methods in two public datasets. We also experiment extensive ablation studies to validate  the impact of each component of the searched model.

\section{Method}
In this section, we first introduce the common neural network architecture adopted by existing works on MRI reconstruction and the residual-in-residual block adapted to the network.
Then, we describe the possible operations inside the block and the detail of how to automatically find the optimal composition of these operations  via the differentiable NAS.

\subsection{Common Network Architecture for MRI Reconstruction}
The goal of MRI reconstruction problem is to learn a function $f$ that maps from undersampled data $x$ to fully-sampled data $y$.
Recent works attempt to approximate the function via a deep neural network, which has achieved promising reconstruction results~\cite{schlemper2017deep,sun2018compressed,zheng2019cascaded}.
Although various deep neural networks have been proposed to increasingly boost the reconstruction performance, we find that most of these networks share the same backbone of neural architectures.
An example of a common architecture is illustrated in Fig.~\ref{fig:basic}. 
The input $x$ is a zero-filled image and output $y$ is a reconstructed image, and we have $x,y\in\mathbb{R}^{2\times w\times h}$, where channel $2$ represents the real and imaginary parts, $w$ and $h$ are the width and height of the image, respectively.
The common architecture consists of $N$ stacked components that have the same \textit{Conv--BasicBlock--Conv--TDC} structure but are optimized separately.
In each component, 
the first convolutional layer \textit{Conv} extracts feature maps of size $c\times w\times h$ from the original input.
Then the \textit{BasicBlock} is a customized operation to further capture deep features of the input and the common choices are sequential convolutional layers~\cite{schlemper2017deep}, recursive dilated block~\cite{sun2018compressed} and dense block~\cite{zheng2019cascaded}. 
The second \textit{Conv} maps $c$ channel features back to the original input size.
The last \textit{TDC} is a two-step data consistency layer \cite{zheng2019cascaded}: (i) replace specific $k$-space value with the original sampled one; (ii) convert the result to real-valued format by calculating its absolute value and then apply step one again. 
The \textit{TDC} layer aims to overcome the inconsistency problem in both $k$-space and image domains.
Let $\mathbf{w}$ be the network parameters, which are usually optimized by a $l_2$ loss function $L(\mathbf{w})=\|y-f(x;\mathbf{w})\|_2^2$.

\begin{figure}[!t]
    \centering
    \includegraphics[width=\linewidth]{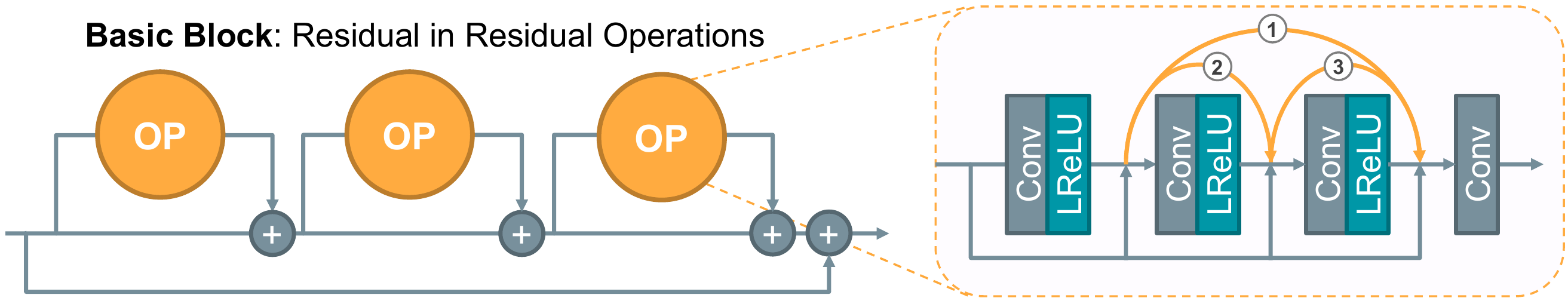}
    \caption{Illustration of residual in residual operations.}
    \label{fig:rir}
\end{figure}

Ideally, deeper neural networks (in terms of both $N$ and depth of \textit{BasicBlock}) are more likely to approximate complex functions and expected to achieve better reconstruction performance.
However, in practice, stacking such components many times may suffer the vanishing gradient problem, which in turn degrades the performance.
Most works in MRI reconstruction adopt skip connection and residual operation in the \textit{BasicBlock} to alleviate the issue of vanishing gradient.
In this work, we adapt the residual-in-residual (RIR) technique \cite{wang2018esrgan} to the \textit{BasicBlock}, which has been recently shown to be effective in very deep neural networks used in super resolution applications.
As shown in Fig.~\ref{fig:rir}, the \textit{RIR BasicBlock} is made of three operation units (OP) and they are sequentially composited via ``residual in residual''.
In each OP, Batch Norm layers (BN) are not placed after Leaky ReLU since existing works have proven such a design can further boost performance and reduce computational complexity in different PSNR-oriented tasks \cite{lim2017enhanced,nah2017deep}.
Meanwhile, multi-level residual dense blocks are adopted, which employ a deeper and more complex structure to learn different level representations, resulting in a higher network capacity~\cite{wang2018esrgan}.

Beyond going deeper, the composition of the network can also become more ``heterogeneous''.
Note that existing works usually adopt the homogeneous structure for all \textit{BasicBlocks}, which may limit the ability of feature representation.
In this work, we attempt to explore if heterogeneous structures of these \textit{BasicBlocks} can further improve the reconstruction performance.
Instead of manually experimenting different compositions of OPs, 
we use differentiable NAS techniques to automatically determine the optimal combinations of cells.

\subsection{Neural Architecture Search for MRI Reconstruction}\label{sec:ss}
We introduce the \textbf{EMR-NAS} (Enhanced MRI Reconstruction Network via neural architecture search) for automatically determining operations in the blocks. 

\noindent\textbf{Search Space}\quad
In NAS, the first step is to design the search space. 
Our search space is based on the OP structure in  Fig.~\ref{fig:rir} (right). 
We introduce eight different cells $\mathcal{O}=\{ O_i, i=1\cdots 8 \}$ that are listed in Table~\ref{tab:search}. 
Specifically, all convolutional layers in OP have the same kernel size of $3\times3$, but different dilation rate and the connection between them.
The dilation rate of each convolutional layer are listed in the second row of Table~\ref{tab:search}. 
The dilation rates of $O_1\ldots,O_7$ are 1-2-4-1, which induce larger receptive fields than that of 1-1-1-1, and are proven to benefit the reconstruction performance~\cite{sun2018compressed,zheng2019cascaded}.
For connection, we only consider the connections numbered with ``1'', ``2'' and ``3'' in Fig.~\ref{fig:rir} (right). 
As the third row in Table~\ref{tab:search} shows that,
$O_1$ and $O_8$ are densely connected. 
$O_2$ to $O_4$ have two connections while $O_5$ to $O_7$ have only one connection.
Our goal is to boost the representation capacity of the network by automatically choosing the best operation within each cell.

\begin{table}
\centering
 \renewcommand\arraystretch{1.1}
\renewcommand{\tabcolsep}{4.5pt}     
\caption{Search Space Design. }
\label{tab:search}
\begin{tabular}{l|c|c|c|c|c|c|c|c}
\toprule
\textbf{Operation} & $O_1$  & $O_2$  & $O_3$ & $O_4$  & $O_5$  & $O_6$  & $O_7$ & $O_8$  \\
\hline
\textbf{Dilation}   & 1,2,4,1 & 1,2,4,1 & 1,2,4,1 & 1,2,4,1 & 1,2,4,1& 1,2,4,1 & 1,2,4,1 & 1,1,1,1 \\
\hline
\textbf{Connection} & 1,2,3       & 2,3         & 1,3         & 1,2           & 1           & 2           & 3         & 1,2,3   \\
\bottomrule
\end{tabular}
\end{table}

\noindent\textbf{Search Strategy}\quad
Suppose there are $T$ OP cells in the network, where every three cells form a \textit{RIR BasicBlock}. 
This results in totally $8^{T}$ different architectures.
Let $O_i^l$ be the $i^{th}$ operation of the $l^{th}$ cell~($i\in[8],l\in[T]$). 
We relax the categorical choice of an operation to the softmax over all possible operations~\cite{liu2018darts}:
\begin{equation}\label{eq:softmax}
x^{l+1} = \sum_{i=1}^8p_i^lO_i^l(x^l)=\sum_{i=1}^8\frac{\exp(\alpha_i^l)}{\sum_{j=1}^8\exp(\alpha_j^l)}O_i^l(x^l),
\end{equation}
where $x^l$ and $x^{l+1}$ denote the input and output of the $l^{th}$ cell respectively. 
The probability $p_i^l$ of choosing the corresponding operation is calculated by the softmax over the architecture parameters $\alpha^l\in\mathbb{R}^{8}$ for the $l^{th}$ cell.

Due to the huge search space, optimizing all the architecture parameters $\mathbf{\alpha}=\{\alpha^1,\ldots,\alpha^T\}$ requires lots of computation and large storage in the memory. 
To save memory and speed up the search process, we follow the path binarization strategy proposed in~\cite{cai2018proxylessnas}.
In particular, the probability $p$ of a specific cell is transformed into binary gates:
\begin{equation}
g = \mathrm{binarize}(p_1, \ldots, p_8)=
\begin{cases}
[1,0,\cdots,0] & \text{if } p_1=\argmax_{i}p_i\\ 
\quad\quad\cdots\\ 
[0,0,\cdots,1] & \text{if } p_8=\argmax_{i}p_i
\end{cases}
\end{equation}
Based on the binary gates $g$, the output of mixed operations of the $l^{th}$ cell is given by $x^{l+1} =\sum_{i=1}^8g_i^lO_i^l(x^l)$.
After binarization of probabilities, only one path is activated in memory at run-time and the memory decreases to the same level of training a single model. 
The relaxation of Eq.~\ref{eq:softmax} makes the network's training and search possible to be differentiable. 
We partition the whole dataset into: $\{\mathcal{S}_{train}, \mathcal{S}_{val}, \mathcal{S}_{test}\}$.
Note that the network weights are optimized on $\mathcal{S}_{train}$ and the architecture parameters $\alpha$ are optimized on $\mathcal{S}_{val}$. 
We search and optimize the network $\mathbf{w}$ in an alternative way, which is given in Algorithm~\ref{alg:search}.
After obtaining the optimal $\alpha$ and discretizing it by $argmax$, we fix the  architecture and retrain the network on the $\mathcal{S}_{trainval} = \{\mathcal{S}_{train}, \mathcal{S}_{val}\}$ and then test it on $\mathcal{S}_{test}$.

\begin{algorithm}[]
\caption{EMR-NAS}
\label{alg:search}
\begin{algorithmic}[1]
\State \textbf{Input:} training set $\mathcal{S}_{train}$, validation set $\mathcal{S}_{val}$, mixed operations $\mathcal{O}$.
\State \textbf{Output:} network weights $\mathbf{w}$, architecture parameters $\alpha$.
\State Warmup the training for $M$ epochs.
\While{training not converged}
    \State // Train $\mathbf{w}$
    \State Reset binary gates by $p=softmax(\alpha)$, active chosen paths.
    \State Update network weights $\mathbf{w}$ by gradient descent $\nabla_{\mathbf{w}}L_{train}(\mathbf{w}, \mathbf{\alpha})$.
    \State // Train $\mathbf{\alpha}$
    \State Reset binary gates by $p=softmax(\alpha)$, active chosen paths.
    \State Update architecture parameters $\mathbf{\alpha}$ by gradient descent $\nabla_{\mathbf{\alpha}}L_{val}(\mathbf{w}, \mathbf{\alpha})$.
\EndWhile
\State\Return $\mathbf{w}$ and $\alpha$
\end{algorithmic}
\end{algorithm}

\section{Experiments}
We evaluate the performance of our proposed model on two public datasets. \\
(1) \textbf{Cardiac.} We use the same short axis cardiac datasets as in work~\cite{zheng2019cascaded}, which is created by the work of Alexander \etal~\cite{andreopoulos2008efficient}. 
Each subject's sequence consists of 20 frames and 8-15 slices along the long axis. 
In all, it contains 4480 cardiac real-valued MR images from 33 subjects. 
The image size is $256\times256$. \\
(2) \textbf{Brain.} The Calgary-Campinas-359 dataset is provided by the work~\cite{souza2018open}. 
It includes 35 fully-sampled subjects of T1-weighted MR, which are acquired on a clinical MR scanner. 
The original raw data are acquired with a 12-channel imaging coil and are reconstructed using vendor supplied tools to make them into a single coil image. 
The matrix size is also $256\times256$.
We use random Cartesian masks with 15\% sampling rate for both datasets.

To fully validate the proposed method, we perform a 3-fold cross-validation in the following experiments. 
One fold $\mathcal{S}_{test}$ is for testing and the remaining two folds are separated into $\mathcal{S}_{train}$ and $\mathcal{S}_{val}$ with a ratio around 9:2.
Three different architectures are achieved due to different $\{\mathcal{S}_{train}, \mathcal{S}_{val}\}$ of each fold. 
We adopt an ensemble method by summing up probabilities $p$ of three folds and discretizing the aggregated probabilities to form the optimal architecture.
We set $N=5$ and $T=3\times N=15$.
The size of search space is $8^{15}$.
The warmup training epochs $M=50$ and the search process takes another 50 epochs.
For training the network parameter $\mathbf{w}$, we use the Adam optimizer with a base learning rate $10^{-3}$ with a cosine annealing schedule, a $0.9$ momentum and weight decay of $10^{-7}$. 
For training the architecture parameters $\alpha$, we also adopt Adam optimizer with a learning rate of $10^{-3}$ and weight decay $10^{-6}$.
All models are trained with a batch size of 8. 
It takes around 0.5 Quadro RTX 8000 GPU day for one fold training.

\subsection{Comparisons to State-of-the-Art}
In this experiment, we show the reconstruction performance of our proposed model compared with some state-of-the-art methods, including UNet~\cite{hyun2018deep}, DCCNN~\cite{schlemper2017deep}, RDN~\cite{sun2018compressed} and CDDNTDC~\cite{zheng2019cascaded}.
For our model, 
the resulting operations in each OP cell after the architecture search is as follows.
We obtain $[O_5~O_8~O_8|O_8~O_8~O_8|O_4~O_1~O_2|O_8~O_8~O_8|O_6~O_8~O_8]$ for Cardiac dataset, and
$[O_6~O_6~O_2|O_4~O_1~O_2|O_3~O_8~O_1|O_3~O_6~O_3|O_3~O_1~O_3]$ for Brain dataset.

We also found that $O_8$ is the most frequent one in the Cardiac dataset, while $O_3$ is the most frequent one in the Brain dataset. 
Dense connections are not optimal in most cases nor the dilated convolution since useless information may be filtered by using fewer connections.
For both datasets, good cell diversity may achieve better results as the searched  architectures are heterogeneous across different RIR BasicBlocks.
In fact, in section~\ref{sec:abla}, the model with repeated blocks underperforms our searched heterogeneous architecture.

\begin{figure}[!t]
    \centering
    \includegraphics[width=\linewidth]{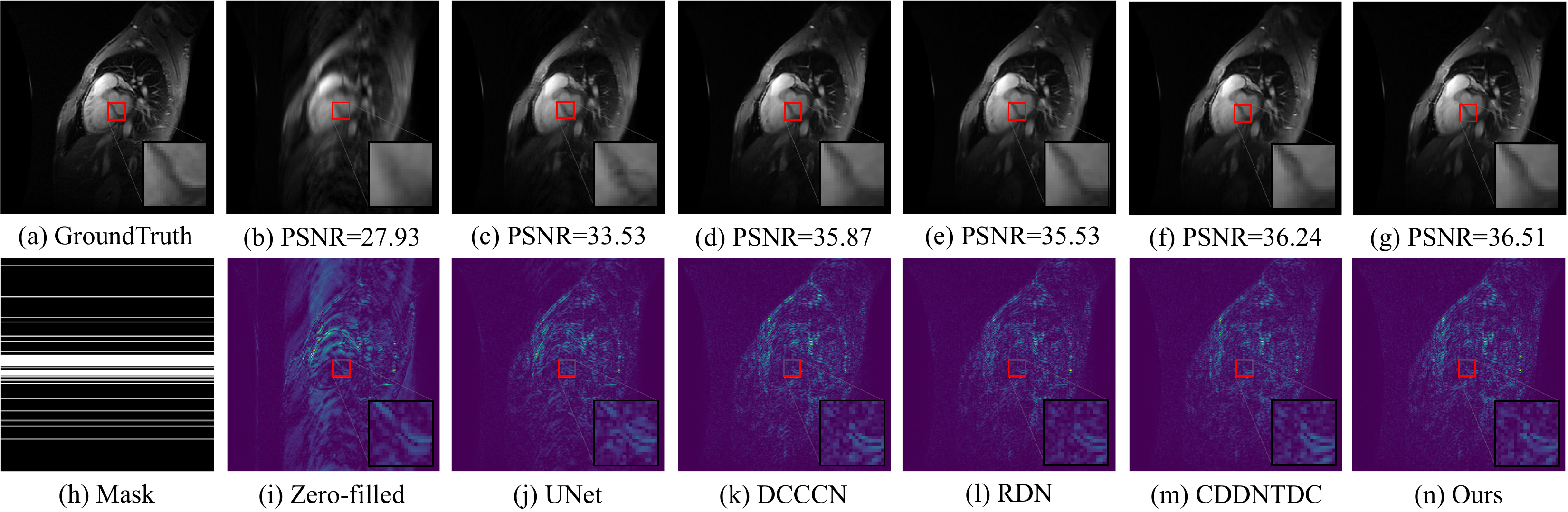}
    \caption{Example of reconstructed images of all methods on the Cardiac dataset.}
    \label{fig:cardiac}
\end{figure}

\begin{figure}[!t]
    \centering
    \includegraphics[width=\linewidth]{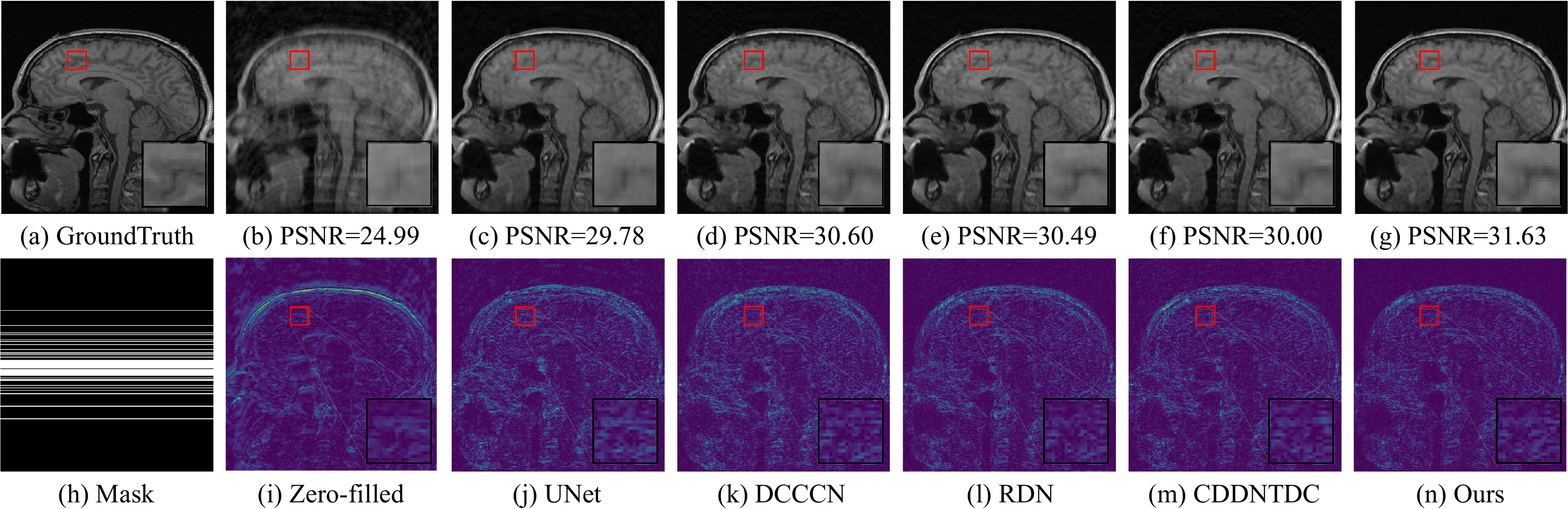}
    \caption{Example of reconstructed images of all methods on the Brain dataset.}
    \label{fig:brain}
\end{figure}

The 3-fold cross-validation test performance of all methods are reported in Table~\ref{tab:SOTA}.
For both datasets, our proposed model achieves the best performance in terms of both PSNR and SSIM.
Especially for the Brain dataset, our approach outperforms other methods by a large margin. 
Example reconstruction results and the corresponding errors are shown in Fig.~\ref{fig:cardiac} and Fig.~\ref{fig:brain}. 
Improvements achieved by our method are highlighted by the red box. 

\begin{table}[t]\label{tab:SOTA}
\centering
\setlength{\tabcolsep}{3.5pt}
\caption{Results of the proposed method on Cardiac and Brain datasets.
The mean$\pm$std values of 3-fold cross validation indices (PSNR and SSIM) are presented.
}
\begin{adjustbox}{width=\linewidth}
\begin{tabular}{llcclcc}
\toprule
\multirow{2}{*}{Model} && \multicolumn{2}{c}{Cardiac} && \multicolumn{2}{c}{Brain}        \\
\cmidrule{3-4} \cmidrule{6-7}
                     && PSNR & SSIM && PSNR & SSIM \\
\midrule
UNet~\cite{hyun2018deep}         && 30.9877$\pm$0.9676 & 0.8516$\pm$0.0110 && 29.8066$\pm$0.0767 & 0.8408$\pm$0.0031 \\
DCCNN~\cite{schlemper2017deep}   && 34.1993$\pm$0.8519 & 0.9235$\pm$0.0025 && 30.9349$\pm$0.2271 & 0.8687$\pm$0.0082 \\
RDN~\cite{sun2018compressed}     && 34.0686$\pm$0.9440 & 0.9224$\pm$0.0055 && 31.1769$\pm$0.4659 & 0.8717$\pm$0.0068 \\
CDDNTDC~\cite{zheng2019cascaded} && 34.4631$\pm$0.9161 & 0.9291$\pm$0.0029 && 29.9225$\pm$0.1042 & 0.8389$\pm$0.0027 \\
Ours                             && \textbf{34.8653$\pm$0.9126} & \textbf{0.9342$\pm$0.0028} && \textbf{31.7616$\pm$0.0774} & \textbf{0.8882$\pm$0.0011} \\
\bottomrule
\end{tabular}
\end{adjustbox}
\end{table}

\subsection{Ablation Study}\label{sec:abla}
We study the contributions of different components of our model on the Cardiac dataset. 
Seven variants (column A to G) are designed and listed in Table~\ref{tab:ablation} with six different factors related to our model.
\textit{Batch Norm} or \textit{RIR} indicate if BN layer or residual-in-residual is adopted.
\textit{Search} represents whether the architecture is searched by the strategy in Section~\ref{sec:ss} or randomly assembled using operations $\mathcal{O}$.
\textit{Ensemble} means if we ensemble three different architectures from 3-folds 
or adopt the architecture searched from the corresponding fold. 
\textit{Deeper} means 4 cells in each \textit{BasicBlock} rather than the original 3 cells.
\textit{Homogeneous} means if the \textit{BasicBlocks} consist of single operation or multiple opearations.

\begin{figure}[!t]
\centering
\includegraphics[width=0.485\linewidth]{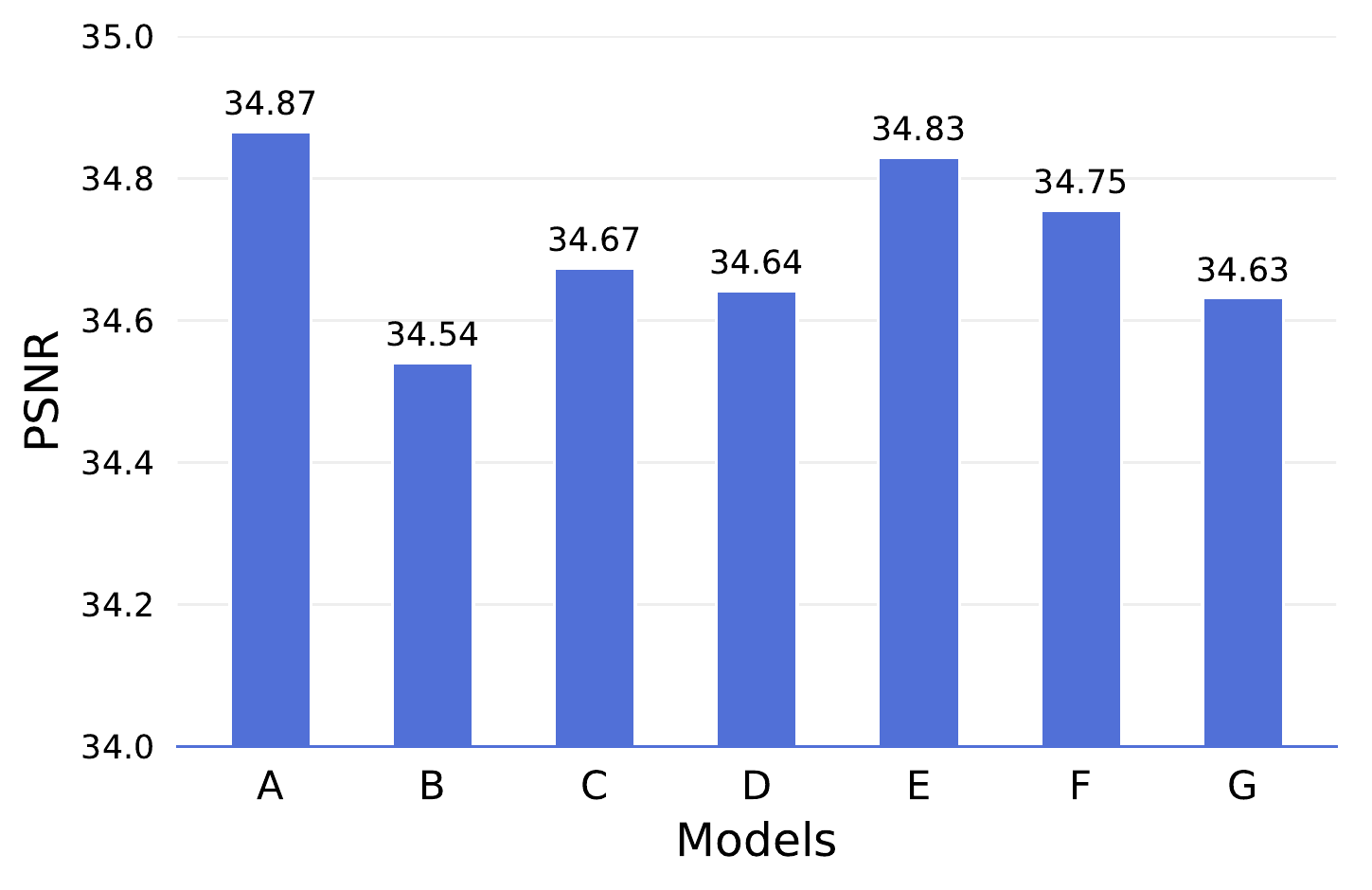}
\hfill
\includegraphics[width=0.485\linewidth]{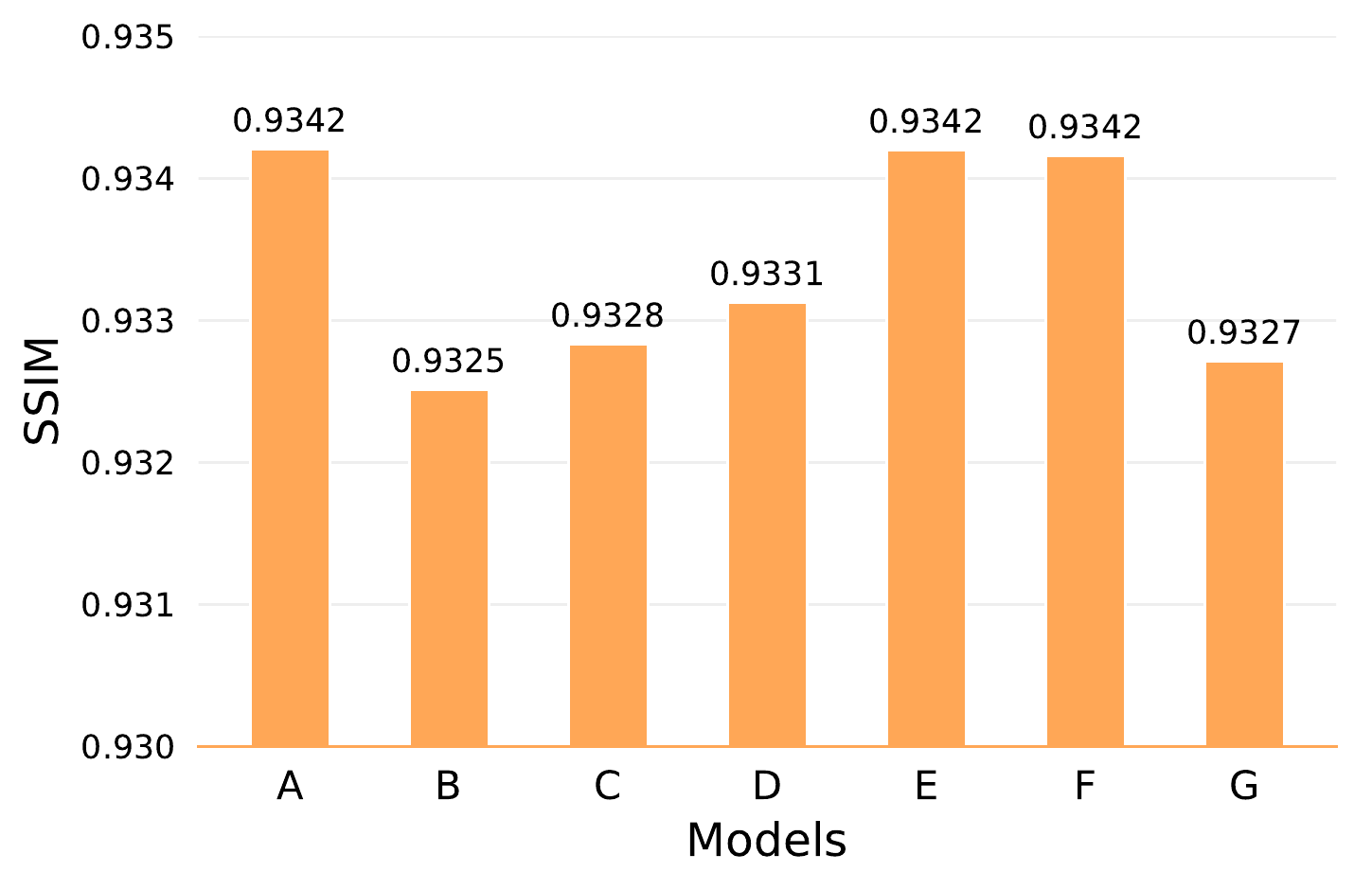} \\
\caption{The PSNR (left) and SSIM (right) results of ablation studies.}
\label{fig:results}
\end{figure}

The PSNR and SSIM results of seven variants are plotted in Fig.~\ref{fig:results}. 
Basically, we see that the optimal combination is given by the default model A, whose structure is searched by NAS and equipped with RIR and ensemble.
RIR improves deep network training and NAS helps improve the representation capacity of the network.
In addition, we observe that the ensemble (E) and the deeper architectures (F) play a less important role in the reconstruction performance, since the corresponding scores are very close to the one by model (A).
\begin{table}[!t]
\centering 
\setlength{\tabcolsep}{15pt}
\caption{Contribution of each component in
Our model. 
Model A is the proposed combination and the contradicted component in other variants is marked in red.
}
\label{tab:ablation}
\begin{adjustbox}{width=\linewidth}
\begin{tabular}{lccccccc}
\toprule
Component & A   & B   & C   & D   & E   & F   & G \\
\midrule
Batch Norm?       & No  & \textcolor{red}{Yes} & No  & No  & No  & No & No \\
RIR?      & Yes & Yes & \textcolor{red}{No}  & Yes & Yes & Yes & Yes \\
Search?   & Yes & Yes & Yes & \textcolor{red}{No}  & Yes & Yes & Yes \\
Ensemble? & Yes & Yes & Yes & Yes & \textcolor{red}{No}  & Yes & Yes\\
Deeper?   & No  & No  & No  & No  & No  & \textcolor{red}{Yes} & No\\
Homogeneous? & No  & No  & No  & No  & No  & Yes & \textcolor{red}{Yes} \\
\bottomrule
\end{tabular}
\end{adjustbox}
\end{table}

\subsection{Discussion on Model Size and Efficiency}
For the number of parameters, we observed that U-Net is the largest model with 1.57M parameters, but it achieves the worst performance. 
Our model achieves the best performance using much fewer parameters (0.33M). 
This shows higher model complexity does not always lead to better performance. 
The key is how to design an effective architecture, and the NAS technique can automate this process.
Other models like DCCNN, RDN and CDDNTDC originally use fewer parameters, however, for a fair comparison, we increased their parameters to around 0.33M (same as ours) by adding convolutional layers between each DC layer. 
Note that our method still dominates others over PSNR and SSIM on two datasets, for example, our model achieves 34.865 in PSNR while DCCNN, RDN and CDDNTDC gain 34.301, 33.257 and 33.981, respectively, which means the searched architecture is better. The RIR structure also helps to prevent the gradient vanishing problem when the network goes deep while other methods’ performance is improved when we increase the capacity of the network.
Besides, our model costs 310s per epoch in training and 0.055 s/frame in inference, which is faster than RDN (370s, 0.060s) and CDDNTDC models (441s, 0.065s), but slower than the U-Net (280s, 0.036s) and DCCNN models (288s, 0.026s).

\section{Conclusion}
In this work, we present an enhanced MRI reconstruction network using NAS technique. 
In particular, we use the residual in residual structure as the basic block and design eight different choices in each block.
An automatic differentiable search technique is used to decide the optimal composition of operations. 
We conduct extensive experiments to compare our methods to the state-of-the-art methods and also perform an ablation study to prove the importance of each component of the proposed method. 
The results show the superior performance of our proposed method and the effectiveness of our architectures design.

\bibliographystyle{splncs04}
\bibliography{mybibliography}

\end{document}